
%

%
%
\input  PHYZZX
\nopubblock \vskip0.2in
 \line{\hfill May 1992}
 \line{\hfill\caps UPR--506--T}

\title{Cosmological Implications of
Domain Walls due to Duality Invariant Moduli
 Sector of Superstring Vacua}
\author{M. Cveti\v c and R. L. Davis
}
\address{
Department of Physics\break
University of Pennsylvania\break
Philadelphia, PA 19104--6396\break}
\vfill
\singlespace
\vsize=630pt
\abstract{
We study cosmological implications of the
duality ($PSL(2,{\bf Z})$) invariant potential for the
compactification  radius $T$,
arising  in a class of superstring vacua.
We show that in spite of having
only one minimum in the fundamental domain of the $T$ field
there are two types of non-supersymmetric
domain walls: one is associated
with the discrete Peccei-Quinn symmetry $T\to T+i$,
analogous to the axionic domain wall, and another one
 associated with the noncompact symmetry $T\to 1/T$,
analogous to the $Z_2$ domain walls.
The first one is bound by stringy cosmic strings.
The scale
of such domain walls is governed  by the scale of gaugino condensation
(${\cal O} (10^{16}$ GeV) in the case of hidden $E_8$
gauge group), while the separation between
minima is of order $M_{pl}$.
We discuss the formation of walls and their cosmological implications:
the walls must be gotten rid of, either by
chopping by stringy cosmic strings and/or  inflation.  Since
there is no usual Kibble mechanism to create strings,
 either one must assume
they exist $ab\ initio$, or one must
conclude that string cosmologies require inflation.
The non-perturbative potential dealt with here appears
not to give the needed inflationary
epoch.
\endpage
\unnumberedchapters
     {\it 1. Introduction.}
In $(2,2)$ string compactifications, where $(2,2)$ stands for $N=2$
left-moving as well as $N=2$ right-moving world-sheet supersymmetry,
the vacuum is supersymmetric and contains
a set of massless fields -- moduli $T_i$ -- which have no potential,
{\it i.e.} \REF\Dine{M.~Dine and N.~Seiberg, Nucl. Phys. {\bf B301},
357(1988).}
\REF\Derend{J.~P.~Derendinger, L.~E.~Ib\'a\~nez, and
    H.~P.~Nilles, Phys. Lett. {\bf 155B}, 65 (1985); M.~Dine, R.~Rohm,
    N.~Seiberg, and E.~Witten, Phys. Lett. {\bf 156B}, 55 (1985).}
\REF\Rey{S.-J.~Rey, {\it Axionic String Instantons and Their
    Low-Energy Implications,} Invited Talk at Tuscaloosa Workshop on
    Particle Theory and Superstrings, ed. L.~Clavelli and B.~Harm, World
  Scientific Pub., (November, 1989); Phys. Rev. D {\bf 43}, 526 (1991).}
\REF\Ferrara{S.~Ferrara, D.~L\"ust, A.~Shapere, and S.~Theisen,
    Phys. Lett {\bf 225B}, 363 (1989).}\
\REF\Cvetic{M.~Cveti\v c, A.~Font, L.~E.~Ib\'a\~nez, D.~L\"ust,
    and F.~Quevedo, Nucl. Phys. {\bf B361}, 194
     (1991).    }
\REF\Kaplun{V.~Kaplunovsky, Nucl. Phys. {\bf B307}, 145 (1988).}
\REF\Dixon{L.~Dixon, V.~Kaplunovsky, and J.~Louis,
Nucl. Phys. {\bf B355}, 649 (1991); J.~Louis,
{\it PASCOS 1991 Proceedings}, P. Nath ed.,
World Scientific  1991.}
\REF\FONTII{A.~Font, L.~E.~Ib\'a\~nez, D.~L\"ust, and F.~Quevedo,
    Phys. Lett. {\bf 245B}, 401 (1990)}
\REF\Font{S.~Ferrara, N.~Magnoli, T.~R.~Taylor, and
    G.~Veneziano, Phys. Lett. {\bf 245B}, 409 (1990);
           P.~Binetruy and M.~K.~Gaillard, Phys. Lett. {\bf
    253B}, 119 (1991);
            H.~P.~Nilles and M.~Olechowski, Phys. Lett. {\bf
    248B}, 268 (1990).}
\REF\CV{M.~Cveti\v c, S.~J.~Rey and F.~Quevedo,
Phys. Rev. Lett. {\bf 67}, 1836 (1991).}
\REF\CGR{M. Cveti\v c, S. Griffies and S.-J. Rey, {\it
Static Domain Walls in $N=1$ Supergravity},
UPR-474-T, YCTP-P43-91 (January 1992), Nucl. Phys. {\bf B}, in press.}
$V(T_i)\equiv0$, to all orders in string loops\refmark{\Dine}\ .
Thus, perturbatively, there is a large degeneracy of string vacua, since
any vacuum expectation value of moduli corresponds to the vacuum
solution.  On the other hand it is known that non-perturbative stringy
effects, like gaugino condensation\refmark{\Derend}\ and axionic string
instantons \refmark\Rey ,
give rise to the non-perturbative superpotential.

In the case of the modulus $T$ associated with the internal size of the
compactified space for the so-called flat background compactifications
({\it e.g.}, orbifolds, self-dual lattice constructions, fermionic
constructions) the generalized target space duality is characterized by
non-compact discrete group
$PSL(2,{\bf Z})=SL(2,{\bf Z})/Z_2$ specified by
$$T\rightarrow{{aT-ib}\over{icT+d}}  \ \ \ ,
 ad-bc=1\ \ ,
 \{a, b, c, d\}\in {\bf Z}.\eqn\modtrans
 $$
Note, the dimensionless field $T$ is written as
 $T= R^2/\alpha '+ iA$, where  $R$ is the radius
of compactification, A corresponds to the
internal axion field and $\alpha' (= 32\pi/g^2M_{pl}^2)$
is the string tension. Here, $g$ the gauge coupling (as defined in
grand unified theories) and $M_{pl}$  is the Planck mass.

If one assumes that the generalized target space duality is
preserved even
non-perturba-tively\refmark{\Ferrara,\Cvetic},
the form of the non perturbative superpotential is very
restrictive\refmark{\Cvetic}\ .
The fact that this is an exact symmetry
of string theory even at the level of non-perturbative effects is
supported by genus-one threshold
calculations\refmark{\Kaplun,\Dixon},
which in turn specify the form of the gaugino
condensate\refmark{\FONTII ,\Font}, and thus restrict the
 form of the non-perturbatively induced potential
to a  specific, restricted  form  of the duality invariant potential.
\REF\Kim{J.~E.~Kim, Phys. Rev. Lett. {\bf 43}, 103 (1979);
    M.~Dine, W.~Fischler, and M.~Srednicki, Phys. Lett. {\bf 104B}, 199
    (1981); and Nucl. Phys. {\bf B189}, 575 (1981); M.~B.~Wise,
    H.~Georgi, and S.~L.~Glashow, Phys. Rev. Lett. {\bf 47}, 402 (1981);
    A good review is by J.~E.~Kim, Phys. Rep. {\bf 150}, 1 (1987).}
\REF\Sikivie{P.~Sikivie, Phys. Rev. Lett. {\bf 48}, 1156
    (1982);  G. Lazarides and Q. Shafi, Phys. Lett.{\bf 115B} (1982) 21;
    For a review, see A.~Vilenkin, Phys. Rep. {\bf 121}, 263
    (1985).}
In the case of  the general duality
invariant potentials, \ie ,  invariant
under non-compact discrete symmetry $PSL(2,{\bf Z})$ ,
this leads in general to stable supersymmetric domain
walls\refmark{\CV ,\CGR}.
\REF\CG{M.Cveti\v c and S.Griffies,
{\it  Gravitational Effects in Supersymmetric  Domain
Wall Backgrounds}, UPR-503-T (April 1992), Phys. Lett. {\bf B}, in press.
}
Such domain walls correspond to a new class of domain walls
interpolating  between non-degenerate minima of the
matter potential\refmark\CGR with
intriguing gravitational effects\refmark\CG .

In this note we would like to address the cosmological
consequences of a specific type  of duality
invariant potential,\ie , the one  which is
due to the gaugino condensation\refmark{\FONTII ,\Font}
in the hidden sector  of string theory. In the orbifold-type
compactifications, due  to the
threshold corrections\refmark{\Kaplun\Dixon}, such a duality invariant
potential is of a very specific type\foot{
Even when taking
into account the Green-Schwartz mechanism to
\REF\CO{
G. Cardoso and B. Ovrut, Nucl. Phys. {\bf 369}, 351 (1992),
 Proceedings  of Strings and Symmetries, 1991,
Stony Brook, World Scientific (P. Van Nieuwenhuisen et al. eds.)
and {\it Coordinate and K\" ahler $\sigma$-Model Anomalies and their
Cancellation in the String Effective Field Theories}, UPR-502-T
(May 1992).}
\REF\DER{J.-P. Derendinger, S. Ferrara, C. Kounnas and
F. Zwirner, CERN preprint CERN-TH-6004/91-REV and Phys. Lett. {\bf
 271B}, 307 (1991).}
cancel\refmark{\CO ,\DER}
anomalies of the underlying non-linear $\sigma$ model
 there is in general a duality invariant
\REF\LM{ D. L\" ust and C. Mu\~ nos, {\it Duality Invariant
Gaugino Condensation and One-Loop Corrected K\" ahler  Potentials
in String Theory}, CERN-TH 6358/91 (December 1991).}
\REF\CCM{B. de Carlos, J. Casas and C. Mu\~ noz, {\it
Supersymmetry Breaking and Determination of the
Unification Gauge Coupling Constant in String Theory},
CERN-TH-6346/92 (April 1992).}
 non-zero potential\refmark{\LM ,\CCM}; \eg\ it
appears in the  orbifold compactifications  which have fixed two-tori
in the  twisted sector  of the  theory, like  $Z_4$ orbifold.
We will confine our study to such types of potentials.}.

The interaction of the
modulus field $T$ is described in terms of $N=1$ supergravity
Lagrangian. The field lives in the fundamental domain {\cal D}
determined by the modular group ${\bf C}/PSL(2,{\bf Z})$.
(see fig. 1).
We take the example of a duality invariant potential
associated with one
modulus field $T$   which corresponds to an overall size of the
  $T^6$ torus of compactified space. This is a general enough example
which is to exhibit
 the qualitative features of gaugino induced non-perturbative
potential.
In this case the
  K\" ahler potential and the
superpotential   for the dilaton field $S$ and
the modulus field $T$
can  then  be written in  the form\refmark{\Font ,\LM}
\foot{  In the following   we shall also neglect
corrections due to one-loop corrections.\refmark\DER
}:
$$
\eqalign{
K &= -{1\over \kappa} ln (S+S^*)
-{3\over \kappa}
ln(T + {T}^*)  \cr
W &= \Omega (S)  \eta (T)^{-6}}
\eqn\localstringy$$
with $\kappa\equiv
 8\pi G_N$.
Here, $\eta(T)$ is the Dedekind eta function, a modular form of weight
$1/2$
\REF\modularform{B. Schoeneberg, \sl Elliptic Modular Functions, \rm
Springer, Berlin-Heidelberg (1970);
J. Lehner, \sl Discontinuous Groups and
Automorphic Functions, \rm
 ed. by the American Mathematical Society,
(1964).}
\refmark{\modularform}.
 Note, that
$\eta$ is regular everywhere
in the fundamental domain, and falls off as $e^{-1/12\pi T}$ as $T\to
\infty$.
The coefficient $\Omega (S)$ is in the case of the
one hidden gauge group without matter of the type
$M_{C}^3e^{24\pi ^2/b_0 S}$. $\Omega (S)$
sets the scale of the potential for the $T$ field
and its  exponential dependence on the $S$-dilaton field
indicates that the superpotential is due to the non-perturbative physics.
$b_0$ is related to the one-loop
$N=1$ (negative) beta function $\beta$
 of  the gauge group
responsible  for gaugino condensation as $\beta=b_0g^3/16\pi^2$.
Note, $Re S=1/g^2$ where $g$ is the gauge coupling at the
    scale $M_C =gM_{pl}C$ where, constant
C  depends
on the subtraction scheme used\Ref\VK{V. Kaplunovsky,
Nucl. Phys. {\bf 307B}, 145 (1988).}.  In ${\overline{DR}}$ scheme
C=0.043.
 For $E_8$ hidden gauge group the scale $\Omega (S)^{1/3}$ is
 \REF\Kras{ N.~Krasnikov, Phys. Lett.  {\bf193B}, 37 (1987); L.~Dixon,
V.~Kaplunovsky, and M.~Peskin, unpublished; L.~Dixon, Proceedings of Rice
Meeting of the APS DPF B.~Bonner and H.~Miettinen, eds. (World Scientific,
Singapore, 1990); J.~Casas, Z.~Lalak, Mu\~noz, and G.~Ross, Nucl. Phys. B{\bf
347}, 243 (1990).}
 ${\cal O}(10^{16})$GeV.\foot{In string theory
the origin of a mechanism that fixes the dilaton field
S is not known. One possiblity is the
 multiple gaugino condensation\refmark{\Kras , \CCM}.
 We assume that
$S$ is fixed to be $1/g^2$, with the gauge constant
$g$ having perturbative value $<{\cal O}(1)$ which in turn
determines  scale associated with  $\Omega (S)$.
In  further discussion  we fix the value of $S$ and
suppress  the dynamics associated with the $S$
field, including the issue of supersymmetry breaking in this sector.}

The scalar Lagrangian is of the form\foot{We use
  $(+,-,-,-)$ space-time signature.} :
$$e^{-1}L = -{1 \over {2\kappa}}
 R + K_{T \bar T}g^{\mu \nu}\partial_{\mu}\bar T
\partial_{\nu}T - e^{\kappa
K}(K^{T \bar T}|D_{T}W|^{2} - 3\kappa
|W|^{2})
\eqn\localL$$
where
$e = |detg_{\mu \nu}|^{1 \over 2}$, and
 $D_{T}W \equiv e^{-\kappa K} (\partial_T
 e^{\kappa K} W)$.
For the K\" ahler potential and superpotential in \localstringy
the Lagrangian\localL   (with the  fixed value of $S$)
takes the form
$$e^{-1}L = -{1 \over {2\kappa}}R + {{3g^{\mu \nu}\partial_{\mu} T
\partial_{\nu}T^*}
\over {\kappa (T+T^*)^2}}$$$$-
{{\kappa |\Omega (S)|^2}
\over{(T+{T}^*)^3(S+ S^*)
|\eta(T)|^{12} }}
\left[{{(T+    {T}^*)^{2}}\over 3}
 |{3 \over 2 \pi}\hat{G}_{2}(T,T^*)|^{2}
 -3
  \right].
\eqn\scalarpotential$$
where $\hat G_2=-{4\pi}\partial_T \eta/{\eta} -2\pi/(T+ T^*)$ is the
 Eisenstein function of weight 2\refmark{\modularform} .

Note that without superpotential the moduli sector has a
continuous non-compact symmetry $SU(1,1)\equiv SL(2,{\bf R})$.
The non-perturbatively induced superpotential in
\scalarpotential\
breaks the continuous symmetry down to its maximal
discrete subgroup $PSL(2,{\bf Z})$.
\REF\Giveon{ A. Giveon, E. Rabinovici and G. Veneziano,
Nucl. Phys. {\bf B322}, 167 (1989).}
The potential (see Fig. 3)
in \scalarpotential\ has {\it one} minimum
in the fundamental domain at $T=1.2$ which at the
same time breaks supersymmetry. This turns out
 to be a generic  property\refmark{\Font , \CCM}
for  a  gaugino induced non-perturbative
potential in string vacua with $PSL(2,{\bf Z})$ invariance.

{\it 2.  Domain Walls Connected to Stringy Cosmic Strings.}
 The  underlying $PSL(2,{\bf Z})$ symmetry of the theory
 implies \refmark{\Giveon ,\FONTII ,\CV
  }
 that there could be domain wall solutions
interpolating between such degenerate vacua of different
fundamental domains.

The first class of domain walls is associated with the  symmetry
transformation $T\to T+i$, \ie\   the discrete Peccei-Quinn symmetry. Thus the
domain walls interpolates between minima with $T=1.2 $ and $T=1.2+i$.
Note, that this transformation
is a genuine symmetry of the underlying sting vacua; \eg , the states
of the underlying string vacua remain intact under such a symmetry
transformation.

The nature of this  type of domain wall
 is closely related to the domain wall that
exists for the  QCD  induced potential of the Peccei-Quinn axion $\theta$ when
there is only one quark flavor. In this  case  \ie\
$V=\Lambda_{QCD}^4(1+\cos\theta)$. In general, spontaneously broken global
$U(1)$ Peccei-Quinn symmetry is non-linearly realized through a
pseudo-Goldstone boson, the invisible  axion $\theta =\{0, 2\pi\}$.
Non-perturbative QCD effects  through the axial anomaly  break explicitly
$U(1)$
symmetry down to $Z_{N_f}$, by generating an  effective potential proportional
to $1+\cos N_f\theta$.  This potential  leads to domain wall solutions
\refmark{\Sikivie}\  with $N_f$ walls meeting at the axionic
strings\refmark{\Kim}. In the case of $N_f=1$ there is still a   domain wall,
interpolating between $\theta =0$ and $\theta=2\pi$. The  domain wall  is
bounded by an axionic string which emerged at the first stage of symmetry
breaking of the global $U(1)$ Peccei-Quinn symmetry.

Very analogous situation is taking place in our case; the role of the   stringy
axion field is played by the imaginary part of the $T$ field;  the domain wall
interpolates between $T=1.2$ and $T=1.2 + i$; \ie\
inspite of only one minimum in the fundamental domain, there
is still a domain wall just like in the case of
axionic domain walls.

The analogy with axionic domain walls
can be carried ever further.
Such a stringy
 domain wall  \REF\VAFA{B.
Greene, A. Shapere, C. Vafa and S.-T. Yau, Nucl. Phys. \bf B337 \rm , 1
 (1990).}
is   bound by a type of stringy cosmic strings\refmark{\VAFA}.\foot{
In the case of supersymmetric stringy domain walls\refmark\CV\ ,
which appear in a class of  general  $PSL(2,{\bf Z})$
invariant potentials, the whole configuration of the domain wall
bounded by the stringy cosmic string preserves supersymmetry.}

 They can  form
at the  first stage, when the {\it continuous} non-compact symmetry $SL(2,{\bf
R})$ is spontaneously broken,
\ie\ at this stage there is only the
kinetic energy term present  while
 the non-perturbative potential due to
 gaugino condensation is not been turned on, yet (see
eq.\scalarpotential).

In Ref.\VAFA\ two types   of such stringy  cosmic strings were classified.
They are obtained by mapping the $T$ field configuration on the $(x,y)$ spatial
coordinates  as: $j(T(x+iy))= \sum_na_{n=0}^{n=n_{max}}[(x+iy)/
\sqrt{\kappa}]^n$, where
$n$ are
non-negative integers, $a_n$ are arbitrary constants (presumably
of order one), and $j$ is modularly
invariant function\refmark\modularform .
Note, that $j\to e^{2\pi T}$ as $T\to\infty$, thus  at the
spatial infinity  ($x+iy\equiv r\ e^{i\phi}$ with $r\to \infty$) the $2\pi ImT$
 is mapped into the spatial angle $2\phi n_{max}$ and $Re T$ blows up, \ie\
decompactification takes place. Thus, as $T\to T+i$, the deficit angle is
$\phi=2\pi/n_{max}$. In addition, $j$ function has a triple   zero at
$T=e^{i\pi/6}$ and  near $T=1$ has an
 expansion $j=1728+a(T-1)^2$\refmark\modularform . This
implies, that  such a string has  a core-like structure at two points,  which
are in the region $r={\cal O}(\sqrt\kappa)$,
\ie\  they correspond to  $T=1$
and
$T=e^{i\pi/6}$\refmark\VAFA .
 The second type of domain walls corresponds to
the map $j(T(x+iy))= \sum_na_{n=-n_{max} 0}^{n=0}[(x+iy)/\sqrt{\kappa}]
^n$. In this case the
 string core structure is at spatial infinity and  decompactification takes
place    at $r={\cal O}(\sqrt{\kappa})$.
Note that  the type of stringy domain walls associated with the symmetry
transformation $T\to T+i$ can be bound only by the
first type of stringy cosmic strings (see Fig. 3).

The second type of domain walls are associated with $T\to 1/T$, \ie\ the
generator of the non-compact symmetry  transformation of $PSL(2,{\bf Z})$. This
domain wall interpolates between the minimum at $T=1.2$ and $T=1/1.2$. It is
analogous to the domain walls associated with $Z_2$ symmetry. It is at first
puzzling that there would be such a domain wall, after all, the points in the
$T$ plane are related by the $T\to 1/T$ symmetry. However,
points associated with $T\to  1/T$ transformation
{\it can} be probed since they correspond to a different theory (with heavy
winding modes becoming light and vice versa) which happens to be equivalent to
the original theory.

The second type ($T\to 1/T$) of domain walls  are different in nature as
opposed to the first type ($T\to T+i$) of domain walls. In particular, the
second type of domain walls do not seem to be bound by stringy  cosmic strings
of the type described above.

The features of non-perturbatively induced potentials
associated with the non-compact discrete  symmetry
for the radial moduli of
other (2,2) string vacua  are generic.\foot{See,
\eg, the study of properties of non-compact
discrete symmetry for  another string vacuum\Ref\COP{ P. Candelas, X.
De la Ossa, P. Green and,
L. Parkes,  Phys. Lett. {\bf  258B}, 118 (1991).}.}
In general, the non-compact discrete symmetry group associated
with the radial modulus field has two types of symmetry
generators: there is an  analog of the discrete Peccei-Quinn symmetry, \ie\
$T_k\to T_k +c$, with c being an integer   or half integer, as well as  an
analog of the non-compact symmetry (associated with the symmetry of small and
large  radius of compactification). Thus,
the two types of domain walls
discussed above seem to be a    generic property of the gaugino induced
non-perturbative potential which respects  the non-compact discrete
symmetry of a particular (2,2) string vacuum and possesses
only one minimum in the fundamental domain.

{\it 3.  Cosmological Implications.}
In an expanding hot big bang cosmology the walls form when the non-perturbative
potential turns on. The scale of the domain walls depends on the scale $\Omega
(S)$, \ie\ the scale at which gaugino condensation takes place. This scale
could be as low as ${\cal{O}}$(TeV)
in the case of multiple gaugino condensations\refmark{\Kras ,\CCM}.
In the case of hidden large gauge  groups without matter
a more plausible scale is
${\cal{O}}(10^{16}\hbox{GeV})$, which
is favored in the case of hidden $E_8$ gauge group.
The domain walls produced by a phase transition such as this would
have degenerate vacua on either side and would
quickly come
to dominate the universe.
Thus, the walls have to go away in order to be consistent with
observations.

In ordinary field theory such domain walls
can disappear
in two ways:  one possibility is inflation\Ref
\inf{A. H. Guth, Phys.Rev. {\bf D23}, 347 (1981);
A. D. Linde, Phys. Lett. {\bf 108B}, 389 (1982);
A. Albrecht and P. J. Steinhardt, Phys. Rev. Lett. {\bf 48},
1220 (1982).},
another is chopping
by a brownian network of cosmic strings that appeared at
an earlier phase transition\Ref\walstrng{T. W. B. Kibble,
 G. Lazarides and
Q. Shafi, Phys. Rev. {\bf D26}, 435 (1982);
A. Vilenkin and A. E. Everett, Phys. Rev. Lett. {\bf 48}, 1867 (1982);
A. E. Everett and A. Vilenkin, Nucl. Phys. {\bf B207}, 43 (1982).}.
It is the second possibility which we will examine in the
remainder of this section.

The difference between the
standard picture and the present case is that the
stringy cosmic strings are inherently stringy;
there is no phase transition in the four dimensional
effective field theory at which
the strings appear.  Rather, if they were to exist they would
have to have come directly from the spontaneous
compactification  at the string epoch above $M_C=0.043\  g\ M_{Pl}$, \ie\
the scale below which the physics of the
effective four dimensional field theory takes over.

If so, then a problem for incorporating stringy
cosmic strings in a cosmological scenario
is that there is apparently no natural choice for the
string configuration upon coming out of the Planck era.
Recall that in cosmic string theories of galaxy formation the strings are
assumed, with some numerical and analytical support, to evolve along a scaling
solution in which the fraction of the energy density in strings to that of the
background is constant.  On scales outside the horizon the network executes a
random brownian walk, its initial conditions set up at the time of a phase
transition according to the usual
Kibble mechanism\Ref\kib{T. W. B. Kibble,
J. Physics {\bf A9},
3320 (1976); Physics Reports {\bf 67}, 183 (1980).}. Note that the
usual
Kibble mechanism only
applies to strings which are created by
a VEV appearing at a phase transition.  Just having fundamental strings in
thermal equilibrium will not
 give a network of strings stretching beyond
the horizon.
In the case of stringy
cosmic strings there is no such phase transition.
We thus have no basis on which we can conclude
that there would be a network of
strings extending beyond the horizon.  This problem is due
to our lack of knowledge of superstring dynamics.

In spite of our ignorance of a mechanism by which they can be produced,
let us assume that there is a Brownian string network
present at the time that the
non-perturbative potential turns on.  The domain walls are then created bounded
by strings. In our case such domain walls are bounded by stringy cosmic strings
of  the first type, only. {\it E.g.,}
 $j(T(x+iy))=a_1(x+iy)+a_0$. In this case cosmic
string maps   $2\pi Im T\to\phi$  as  $r\to\infty$. This is thus the type of a
string that is attached to one end of the large domain wall. The energy per
unit length of this string is\refmark\VAFA\ in our case:
$\mu = 3/\kappa\times 2\pi/12$.
In addition, the
core of such a string is more complicated as discussed above (see Fig. 3).
Namely, the domain  wall is bound by a
string with the core like structure
corresponding  to two points:
the $Z_3$ symmetric point at  $T=e^{i\pi/6}$
and $Z_2$ symmetric point $T=1$. Also, the thickness
of the domain wall
(${\cal O} ([\kappa |\Omega (S)|^2]^{-1/4})$)
can be much larger compared to the size of the string
(${\cal O}(\sqrt\kappa)$).

Now suppose that walls form at a scale of $\sigma\sim
{\cal O}
(10^{15})$GeV, where $\sigma$ is the energy per unit area of the wall.
We assume
that a typical radius of curvature of a string at time $t$
is less than $t$,
as is the case for standard cosmic strings.  Then by the time
$t\sim(\mu/\sigma)$, where $\mu$ is the string
energy per unit length, the walls will dominate
the dynamics of the system.  However, because of the string
network there will be no infinite domain walls.  The
walls-bounded-by-string quickly vanish by chopping
into ever smaller pieces.  Even if we make the extreme assumption
that they fall into periodic motion, their lifetimes are limited
by gravitational radiation to be less than $\tau\sim1/\sigma$, and
the walls never come to dominate the universe\Ref\vil{A. Vilenkin,
Phys. Rep. {\bf 121}, 263 (1985).}.

3. {\it Inflation?}
If the absence of a standard Kibble
mechanism to explain the existence of an initial
string network means that the walls form unconnected to
strings, then the only way to eliminate them is inflation. The non-perturbative
potential in\scalarpotential\ for the $T$ field could in principle
allow\Ref\CQRII{ M. Cveti\v c, F. Quevedo and S.-J. Rey, unpublished.} for the
inflationary epoch of the $T$ field. Namely the potential shares common
features with the pseudo-Goldstone potential $\Lambda^4[\cos (\phi/f)+1]$ with
$\Lambda={\cal O} (10^{16})$GeV and $f={\cal O}(M_{pl})$, which was
shown\Ref\FF{ K. Freese, J. Frieman and A. Olinto, Phys. Rev. Lett. {\bf 65},
3233  (1990).} \REF\OT{B. Ovrut and S. Thomas, Phys. Lett. {\bf  267B},
227 (1991) and {\it
Instantons in Antisymmetric Tensor Theories in Four-Dimensions},
 UPR-0465-T (1991).}
to account for a version
of inflation which is argued to be less finely-tuned than others .\foot{ Within
the study of instantons  in a theory with the antisymmetric tensor, such
constraint can also be met\refmark\OT.} In the case when $f\geq M_{pl}$,
inflation takes place for a  wide range of  intial field configurations, \ie\
for  $\phi\to[0,{\cal O}(1)]$, while at the same time producing sufficient
density fluctuations $\delta \rho/\rho={\cal O} (10^{-4})$.

In our case,  the  major difference is that the field $T$ is complex, and thus
it rolls in the two-dimensional space. However,  the potential is close to
satisfying the  two basic constraints. First, its overall scale is governed by
the scale  of gaugino condensation. From \scalarpotential\ one obtains $\Lambda
= [\kappa |\Omega (S)|^2]^{1/4}$,  which in the case of hidden $E_8$
gauge group  corresponds to the energy scale $10^{15}$GeV. In order to
determine the scale $f$, which governs the rate at which the potential changes,
one has to normalize the kinetic energy in \scalarpotential ; this implies that
the  dimensionful field $\phi$ is related to the dimensionless field $T$ as
$\phi=T/\sqrt\kappa$. On the other hand  the adjacent minima,  are separated by
a change in $T$ of order one,  \eg\ along the imaginary direction $T=1.2$ and
$T=1.2 +i$. This in turn implies that the scale
  $ f={\cal O}(1/(2\pi \sqrt\kappa))$.

In the  fundamental domain $\cal D$ a  variation of the potential from its
maxima and saddle points to its global minimum takes place  for the region $Im
T=[0,1/2]$ and   $Re T= [1, 1.2]$ (see Fig. 2).  In our case, although one
naturally expects $f\sim M_{pl}$, when the field is properly rescaled we find
that $f$ is an order of magnitude too small for inflation to proceed via this
truncation of the full theory. We must therefore have inflation arise from
dynamics not considered in the present article. Another problem  of the
concrete example based on $PSL(2,{\bf Z})$ symmetry is the fact that the
supergravity potential in \scalarpotential\ has a negative cosmological
constant. However, the issue of ensuring zero cosmological constant is a
question hardly addressed within any basic theory.

To conclude we would like to stress again that certain features  of
non-perturbatively induced potentials for the radial moduli of other (2,2)
string vacua are generic.  In general,   the non-compact  discrete symmetry
group has two types of symmetry generators: there is an  analog of the discrete
Peccei-Quinn symmetry, \ie\ $T_k\to T_k +c$, with c being an integer   or half
integer, as well as  an analog of the non-compact symmetry (associated with the
symmetry of small and large  radius of compactification). For such moduli the
non-perturbative potential, which preserves the non-compact discrete symmetry
and  possesses only one minimum in the fundamental domain,
in general has the  properties discussed above.  Without a mechanism to
explain the existence of a cosmic
string network we are led to the general conclusion
such superstring universes must have inflation.

We would like to thank S.Griffies, S.-J. Rey and
C. Vafa for useful discussions.
 The work is supported in part by the U.S. DOE Grant
DE--22418--281, junior  faculty SSC Fellowship Award (M.C.),,
and by the NATO Research Grant \#900--700 (M.C.).
\refout
\endpage
$${\bf Figure\ \   Captions}$$

Fig.1  Fundamental domain of the T field.  The dot ($T=e^{i\pi/6}$)
denotes the $Z_3$ symmetric point and the cross $(T=1)$ denotes
the $Z_2$ symmetric point.

FIG.2  The scalar potential for the $T$ field in units of $\kappa
|\Omega(S)|^2$.
  The absolute
minimum lies on the real axis at $T=1.2$.  The figure is taken
from\refmark{\FONTII,\Cvetic}.

FIG.3   Stringy cosmic string connected to a domain wall.  The locations
of the stringy core at $r=0\ (T=e^{i\pi/6})$ and at $re^{i\phi}=1728
\sqrt{\kappa}\
(T=1)$ corresponds to the stringy cosmic string with the map $j(T)=
re^{i\phi}/\sqrt\kappa.$

\end